\documentclass[aps,pre,reprint,amsmath,amssymb]{revtex4-1}
\usepackage{graphicx,subfigure}
\usepackage{dcolumn}
\usepackage{epstopdf}
\usepackage{upgreek}
\usepackage{textcomp}
\usepackage{ulem}
\usepackage{bm}
\usepackage{color}
\newcommand{\bsf}[1]{\textsf{\textbf{#1}}}

\makeatletter
\def\btt#1{\texttt{\@backslashchar#1}}%
\DeclareRobustCommand\bblash{\btt{\@backslashchar}}%
\makeatother

\begin{document}
\title{ Omnidirectional transport and navigation of Janus particles through a nematic liquid crystal film}

\date{\today}
\author{Dinesh Kumar Sahu$^{1}$, Swapnil Kole$^{2}$, Sriram Ramaswamy$^{2}$ and Surajit Dhara$^{1*}$}
\affiliation{$^{1}$School of Physics, University of Hyderabad, Hyderabad 500 046, India\\
$^{2}$Centre for Condensed Matter Theory, Department of Physics, Indian Institute of Science, Bangalore 560 012, India
}

\begin{abstract}

We create controllable active particles in the form of metal-dielectric Janus colloids which acquire motility through a nematic liquid crystal film by transducing the energy of an imposed perpendicular AC electric field. We achieve complete command over trajectories by varying field amplitude and frequency, piloting the colloids at will in the plane spanned by the axes of the particle and the nematic. The underlying mechanism exploits the sensitivity of electro-osmotic flow to the asymmetries of the particle surface and the liquid-crystal defect structure. We present a calculation of the dipolar force density produced by the interplay of the electric field with director anchoring and the contrasting electrostatic boundary conditions on the two hemispheres, that accounts for the dielectric-forward (metal-forward) motion of the colloids due to induced puller (pusher) force dipoles. These findings open unexplored directions for the use of colloids and liquid crystals in controlled transport, assembly and collective dynamics.

 \end{abstract}
\maketitle

Electrophoresis, the use of electric fields to transport tiny particles through fluids, is an important technology for macro-molecular sorting, colloidal assembly and display devices and a challenging area of soft-matter research~\cite{ramos,morgan,drop,stv1,div2,alex,com,rc}. Classic electrophoresis is linear: ions in the electrical double layer drag the fluid, and hence the particle itself, with a velocity proportional to and parallel to the applied field. Induced-charge electroosmosis (ICEO) of particles is nonlinear: the applied field itself creates the double layer. Polarity in the shape or surface properties of the particle results in a flow pattern that picks out a direction of motion, with velocity quadratic in and normal to the field~\cite{murtsovkin,tm,baz1,baz2,baz3}. Neither effect offers the option of continuously tuning the direction of transport and hence the desired motility of the microscopic particles~\cite{sum,wu}.

When the ambient fluid is a nematic liquid crystal (NLC), the anchoring of the mean molecular orientation or director ${\bf \hat n}$ normal to the surface of a suspended homogeneous spherical particle mandates Saturn-ring~\cite{ram,abot} or asymmetric~\cite{lub,stark} defect structures, resulting, respectively, in quadrupolar or dipolar elastic distortions in the NLC~\cite{stark}. The nonlinear electro-osmotic flow resulting from an imposed electric field yields bidirectional transport of dipolar particles parallel to the local director, thanks to their broken fore-aft symmetry, an effect termed LC-enabled electrophoresis (LCEEP)~\cite{od,od1,oleg,oleg1,oleg2}. The Saturn-ring particles, by contrast, maintain the quadrupolar symmetry of the flow and hence display no motility~\cite{oleg1,oleg2}.

Our study focuses on spherical particles with two hemispherical faces, one metal, the other dielectric (see Fig.S1, Supplementary Material~\cite{sup}). Their ``Janus'' character is sensed only by the electrostatics of the medium; as far as the mechanics of the ambient NLC is concerned they are elastic quadrupoles. Our central result is that purely by tuning the amplitude and frequency of an imposed electric field, and not its direction, we can achieve guided transport of Janus colloids in a direction of our choosing perpendicular to the field, amounting to a realization of controllable phoretic active particles \cite{Golestanian_LesHouches}. Our findings suggest novel possibilities at the interface of colloids and liquid crystals for controlled transport, assembly, nonequilibrium phenomena and collective dynamics.\\


\begin{figure}[!ht]
\centering
\includegraphics[scale=1.1]{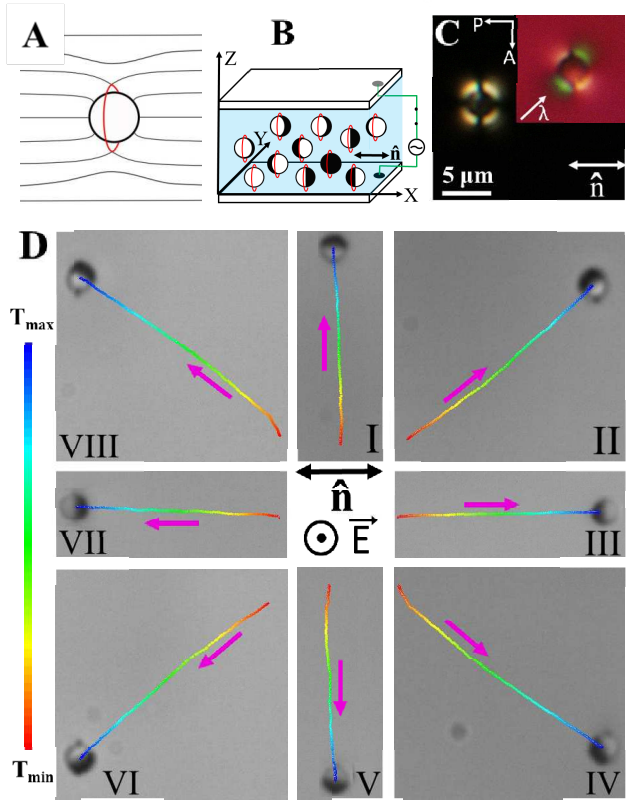}

\caption{ ({\bf A}) Quadrupolar distortion of nematic director around a spherical particle. Red circle represents Saturn-ring defect. ({\bf B}) Janus quadrupolar particles in a planar cell with AC field. The dark hemisphere represents metal. ({\bf C}) Optical microscope texture of a Janus quadrupolar particle in the NLC between crossed polariser (P) and analyser (A) in the $xy$ plane. (Inset) Texture with a $\lambda$-plate (530 nm).  ({\bf D}) Time coded trajectories (T\textsubscript{min} =0s to T\textsubscript{max}=5s ), labelled from I to VIII, under AC electric field (1.54 V$\upmu $m$^{-1}$, 30 Hz). Trajectories of selected particles are grouped.  Pink arrows denote the direction of motion.   (Movie S1 ~\cite{sup}). The direction of electric field, {\bf E} and the director (${\bf \hat n}$) for all trajectories are shown at the center. Cell thickness: 5.2$\upmu$m.}
\label{fig:figure1}
\end{figure}

 We work with a room-temperature nematic liquid crystal mixture MLC-6608. Macroscopic alignment in the $x$ direction is imposed by the treated surfaces of the bounding electrodes parallel to the $xy$ plane (Fig.\ref{fig:figure1}B). Their separation is larger than, but close to, the diameter $2a$ of the suspended particles~\cite{igor}, which produce a quadrupolar elastic field in the nematic (Fig.\ref{fig:figure1}A). We work in the dilute regime and not considering cases of higher concentration where aggregation and network formation are important~\cite{jc,ta}.
Due to the elastic distortion of the director, the particles resist sedimentation and levitate in the bulk~\cite{pis}. This feature, and therefore liquid crystal enabled electrophoresis (LCEEP) as well, are absent in the isotropic phase. The dielectric anisotropy ($\Delta\epsilon=\epsilon_{||}-\epsilon_{\perp}$, where $\epsilon_{||}$ and $\epsilon_{\perp}$ are the dielectric permittivities for the electric field parallel and perpendicular to $\bf\hat{n}$) of the sample is negative so that the electric field ${\bf E}$, applied in the $z$ direction does not influence the macroscopic director except near the particles~\cite{oleg1}. 
Figure \ref{fig:figure1}C shows the optical microscope texture of a Janus quadrupolar particle with cross polarisers. The four-lobed intensity pattern of the particle, a characteristic feature of elastic quadrupole~\cite{igor}, is further substantiated from the texture obtained by inserting a $\lambda$-plate~(inset). The texture without polarisers shows that the metal hemisphere of particles in the absence of  AC field is oriented in different directions, always keeping the Saturn rings perpendicular to the macroscopic director (see Fig.S2A, Supplementary Material~\cite{sup}).  
Depending on the anchoring Janus particles can also induce other types of defects~\cite{mc,mangal}, which are not considered in this study.

Once the AC electric field is switched on, the particles reorient~\cite{sum} so that the plane of the metal-dielectric interface lies parallel to the field (Fig.\ref{fig:figure1}B and Fig.\ref{fig:figure1}D). With increasing field, they start moving in specific directions in the plane of the sample, depending on the orientation of the Janus vector ${\bf\hat{s}}$ (normal to the metal-dielectric interface)  [Movie S1~\cite{sup}]. Real-time trajectories of selected particles are grouped in Fig.\ref{fig:figure1}D. The dielectric hemispheres (Fig.\ref{fig:figure1}D, III and VII) lead when movement is parallel to, and the metal hemisphere (Fig.\ref{fig:figure1}D, I and V) when it is perpendicular to, the macroscopic director. For particles moving at other angles the Janus vector ${\bf\hat{s}}$ interpolates smoothly between these two extremes (Fig.\ref{fig:figure1}D, II, IV, VI and VIII), and the particles can thus move in any direction in the plane of the sample as shown in Fig.S2B (Supplementary Material~\cite{sup}).

\begin{figure}[!ht]
	\begin{center}
	\includegraphics[scale=0.25]{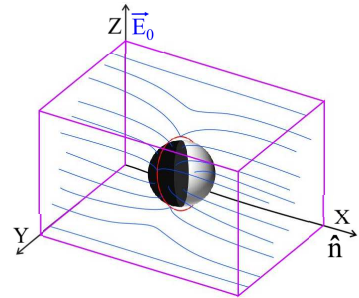}
	\end{center}
\caption{ Quadrupolar director field surrounding a Janus particle. Applied electric field ${\bf E}$ is transverse to the director. Red circle denotes Saturn ring defect. The dark hemisphere represents metal.}
\label{fig:figure2}
\end{figure}

\begin{figure}[!ht]
\begin{center}
\includegraphics[scale=0.4]{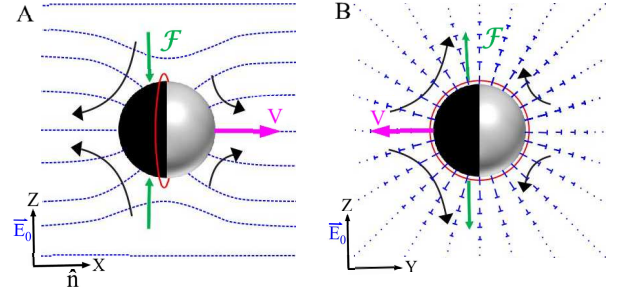}
\end{center}
\caption{ Liquid crystal-enabled electroosmotic flows around a quadrupolar Janus particle in two orthogonal planes parallel to the electric field. Red circles represent Saturn-ring defects. The dark hemisphere represents metal. The large curved arrows on the metal hemisphere indicate stronger flows. The off-centered force dipoles $\boldsymbol{\mathcal{F}}$ (shown in green colour) are ({\bf A}) pullers and ({\bf B}) pushers. The propulsion direction of the particles is indicated by a pink arrow for ({\bf A})  Trajectory III/VII ({\bf B}) Trajectory I/V.}
\label{fig:figure3}
\end{figure}

To understand the motility of a Janus particle in an AC electric field we calculate the electrostatic force density induced by the field in the region around the particle. This force density drives fluid flow over the surface of the particle in such a way as to turn the particle into a swimmer, as sketched in Fig.\ref{fig:figure3}.

We consider a single spherical Janus particle of radius  $a$ in a nematic liquid crystal. The particle's surface consists of a conducting and a dielectric hemisphere, both of which are taken to impose identical uniform homeotropic surface anchoring on the ambient nematic, such as to produce an elastic quadrupolar distortion in the director field, that is, a radial hedgehog compensated by a Saturn ring defect. An electric field ${\bf E_{0}}$ along $z$ is externally imposed on the system as shown in Fig.\ref{fig:figure2}. The mean macroscopic director field lies parallel to the $x$ axis and local deviations from this mean direction are described by an angle $\theta$, positive for counterclockwise rotation. We consider a system with conductivities $\sigma_{||}$, $\sigma_{\perp}$ and dielectric constants $\epsilon_{||}$, $\epsilon_{\perp}$, for electric fields parallel (${||}$) and perpendicular (${\perp}$) to the director $ {\bf \hat{n}}$, and define the anisotropies $\Delta\sigma=\sigma_{||}-\sigma_{\perp}$  and $\Delta\epsilon=\epsilon_{||}-\epsilon_{\perp}$. Our strategy, generalizing \cite{oleg2}, is to use charge conservation and Gauss's Law to obtain the total electric field ${\bf E}$ and the charge density $\rho$ and thus the electrostatic force density $\mathcal{F}({\bf r}) = \rho({\bf r}) {\bf E}({\bf r})$, separately for the case of a dielectric and a conducting sphere and combine these results to infer the character of the induced flow around the Janus sphere. 

For our particles of radius $a \simeq 1.5$ $\upmu$m moving at speed $v \simeq 10$ $\upmu $m/s through a liquid crystal of mass density $\rho \simeq 10^{3}$ kg/m$^{3}$, shear viscosity $\eta \simeq 0.02$ Pa.s and Frank elastic constants $K \simeq 20$ pN (values quoted by supplier Merck KGaA for our sample of MLC-6608), the Reynolds number Re $= \rho v a /\eta \simeq 10^{-7}$ and the Ericksen number Er $= \eta v a /K \simeq 10^{-2}$. We can therefore ignore the effect of fluid inertia and we can take the director configuration around the particles to be negligibly influenced by fluid flow. As in \cite{oleg2}, we work at zero Peclet number, i.e., take the charge currents to be purely Ohmic and not advected by fluid flow. For simplicity we work at low frequencies so that time-dependence can be neglected in the induction equation and thus the electric field ${\bf E} = -\nabla \Psi$ where $\Psi$ is a potential. We will work to first order in the anisotropies $\Delta\sigma$ and $\Delta\epsilon$. We begin by evaluating the potential for a conducting or dielectric particle in the absence of anisotropy, for which the electrostatic boundary conditions imply 
\begin{equation}
\label{polar}
\Psi_0 \equiv -E_{0} z + \lambda E_0\Phi \equiv  - E_{0} z +\lambda \dfrac{E_{0} a^{3} z}{r^{3}}, 
\end{equation} 
\begin{equation}
\label{lambda}
\lambda=\begin{cases}
	1, & \text{conductor},  \\
	\dfrac{\epsilon_{r}-1}{\epsilon_{r}+2} < 1, & \text{dielectric},
\end{cases}
\notag{}
\end{equation}
where $\epsilon_{r}$ is the dielectric constant.

To calculate the induced charge density in the nematic due to anchoring, we impose steady-state charge conservation $\nabla \cdot {\bf J} =0$ for a current ${\bf J} = \boldsymbol{\sigma} \cdot {\bf E} = (\sigma_{\perp} \bsf{I} + \Delta \sigma  {\bf \hat{n}}  {\bf \hat{n}}) \cdot (-\nabla \Psi)$, where $\bsf{I}$ is the unit tensor. This implies 
\begin{equation}
\label{cont}
(\sigma_{\perp} \bsf{I} + \Delta \sigma  {\bf \hat{n}}  {\bf \hat{n}}):\nabla ({\nabla \Psi}) = - \Delta \sigma({\bf \hat{n}} \nabla \cdot  {\bf \hat{n}} +  {\bf \hat{n}} \cdot \nabla  {\bf \hat{n}})\cdot \nabla \Psi, 
\end{equation}
where $:$ denotes contraction with both indices of $\nabla (\nabla \Psi)$. For small deviations $\delta  {\bf \hat{n}}$ about a mean alignment $ {\bf \hat{n}}_0$, and corresponding deviations $\delta {\bf E}$ from a field ${\bf E} = -\nabla \Psi_0$ imposed from the boundaries, writing $\delta {\bf E} = -\nabla  \delta \Psi$ in terms of a potential $\delta \Psi$, \eqref{cont} becomes 
\begin{equation}
\label{contlin}
\begin{split}
\noindent (\sigma_{\perp} \nabla_{\perp}^2 + \sigma_{||} \nabla_{||}^2)\Psi =- \Delta \sigma \big[( {\bf \hat{n}}_0 \nabla \cdot \delta {\bf \hat{n}} +  {\bf \hat{n}}_0 \cdot \nabla \delta {\bf \hat{n}})\cdot  \nabla \Psi_0+ \\ ({\bf \hat{n}}_0  \delta {\bf \hat{n}} +   \delta {\bf \hat{n}}{\bf \hat{n}}_0):\nabla ({\nabla \Psi_0})\big]
\end{split}
\end{equation}
where $||$ and $\perp$ denote components along and transverse to ${\bf \hat{n}}_0$ and $\Psi=\delta \Psi + \Psi_0$. Next, Gauss's Law $\rho = \nabla \cdot {\bf D} = \epsilon_0 \nabla \cdot( \boldsymbol{\epsilon} {\bf E})$ reads, in the same linearized approximation, 
\begin{equation}
\label{Gausslin}
\begin{split}
\rho = -\epsilon_0 (\epsilon_{\perp} \nabla_{\perp}^2 + \epsilon_{||} \nabla_{||}^2)\Psi - \epsilon_0 \Delta \epsilon \big[( {\bf \hat{n}}_0 \nabla \cdot \delta {\bf \hat{n}} + \\ {\bf \hat{n}}_0 \cdot \nabla \delta {\bf \hat{n}})\cdot  \nabla \Psi_0 + ({\bf \hat{n}}_0  \delta {\bf \hat{n}} +   \delta {\bf \hat{n}}{\bf \hat{n}}_0):\nabla ({\nabla \Psi_0})]\end{split}
\end{equation}
Solving \eqref{contlin} for $\Psi$  
 allows us to write the force density  
\begin{equation} 
\label{forcelinear}
\begin{split}
\boldsymbol{\mathcal{F}} = \rho {\bf E} \simeq \epsilon_0 (-\Delta \epsilon+ \Delta \sigma  G^{-1}_{\epsilon} G_{\sigma} ) \big[ ( {\bf \hat{n}}_0 \nabla \cdot \delta {\bf \hat{n}} {\bf \hat{n}}_0  \cdot \nabla \delta {\bf \hat{n}})\cdot \\ \nabla \Psi_0+  ({\bf \hat{n}}_0  \delta {\bf \hat{n}} +   \delta {\bf \hat{n}}{\bf \hat{n}}_0):\nabla ({\nabla \Psi_0})\big] {\bf E}_0 
\end{split}
\end{equation}
where we have defined the Green's functions 
\begin{equation} 
\label{Gsigma}
G_{\sigma} = (\sigma_{\perp} \nabla_{\perp}^2 + \sigma_{||} \nabla_{||}^2)^{-1} 
\end{equation}
and
\begin{equation} 
\label{Gepsilon}
G_{\epsilon} = (\epsilon_{\perp} \nabla_{\perp}^2 + \epsilon_{||} \nabla_{||}^2)^{-1}. 
\end{equation} 
At the lowest order in $\delta {\bf \hat{n}}$ the charge density is driven by the externally imposed electric field ${\bf E}_0$. We show below that the second term (which we shall call $\boldsymbol{\mathcal{F}_{II}}$) in square brackets on the right-hand side of \eqref{forcelinear} contributes only a higher multipole to the force density. The force density thus takes the form 
\begin{equation} 
\label{forceden}
\boldsymbol{\mathcal{F}} \simeq \epsilon_0 (-\Delta \epsilon+ \Delta \sigma  G^{-1}_{\epsilon} G_{\sigma} ) \big[ ( {\bf \hat{n}}_0 \nabla \cdot \delta {\bf \hat{n}} +  {\bf \hat{n}}_0 \cdot \nabla \delta {\bf \hat{n}})\cdot \nabla \Psi_0\big] {\bf E}_0 
\end{equation} 
It is useful to decompose the force density as $\boldsymbol{\mathcal{F}} = \boldsymbol{\mathcal{F}}_0 + \boldsymbol{\mathcal{F}}_{\lambda}$ where $\boldsymbol{\mathcal{F}}_0$ is a contribution independent of whether the sphere is dielectric or conducting, while $\boldsymbol{\mathcal{F}}_{\lambda}$ depends, through $\lambda$, on the electrical nature of the sphere:
\begin{equation}
\label{fo}
	\boldsymbol{\mathcal{F}_0} \simeq -\epsilon_0 (-\Delta \epsilon+ \Delta \sigma  G^{-1}_{\epsilon} G_{\sigma} ) 
	( {\bf \hat{n}}_0 \nabla \cdot \delta {\bf \hat{n}} +  {\bf \hat{n}}_0 \cdot \nabla \delta {\bf \hat{n}})\cdot \hat{\bf z}  E_{0}^2\hat{\bf z} 
	\end{equation}
and 
\begin{equation}
\label{fd}
\boldsymbol{\mathcal{F}_{\lambda}} \simeq \epsilon_0 \lambda (-\Delta \epsilon+ \Delta \sigma  G^{-1}_{\epsilon} G_{\sigma} ) \\
( {\bf \hat{n}}_0 \nabla \cdot \delta {\bf \hat{n}} +  {\bf \hat{n}}_0 \cdot \nabla \delta {\bf \hat{n}})\cdot \nabla \Phi  E_{0}^2\hat{\bf z} 
\end{equation} where $\Phi$ was defined in \eqref{polar} above, and $\boldsymbol{\mathcal{F}}_0$ is the same for both dielectric and conducting surfaces, while the symmetry breaking piece is $\boldsymbol{\mathcal{F}}_{\lambda}$.
From \eqref{fo} and \eqref{fd}, we see the effect is proportional to the square of the electric field and thus survives time-averaging over a period.  Also from \eqref{fd}, the force dipole  is larger for a conducting sphere than for a dielectric sphere by a factor of 1/$\lambda$. Therefore, for a Janus particle, the center of the force dipole shifts from the geometric center towards the conducting side, breaking the symmetry and hence rendering the particle motile in an applied AC electric field. 

\begin{figure*}[!ht]
\centering
\includegraphics[scale=0.8]{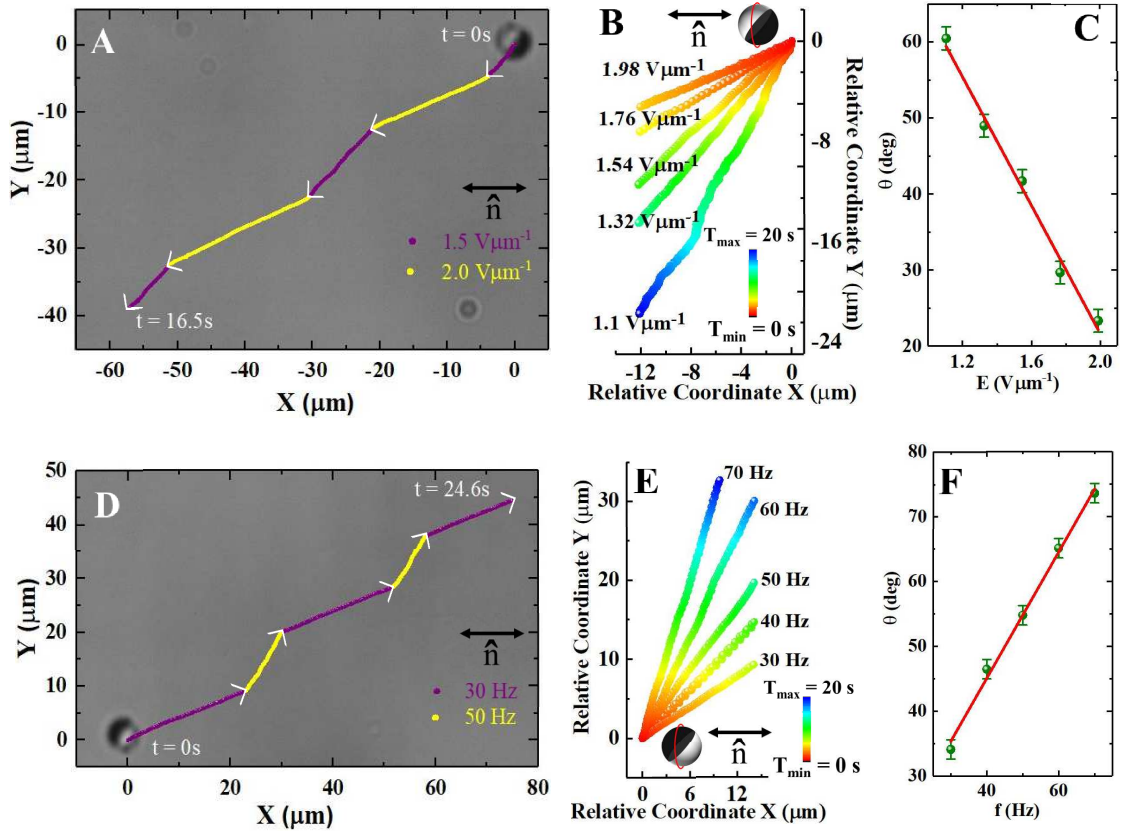}
\caption{({\bf A}) Change of direction of motion of a quadrupolar Janus particle by altering the field amplitude recursively between 1.5 V$\upmu$m$^{-1}$ (purple) to 2.0 V$\upmu$m$^{-1}$ (yellow), keeping the frequency constant at 30 Hz. [Movie S2~\cite{sup}]. Arrows indicate the direction of motion. ({\bf B}) Time-coded trajectories of a particle at different field amplitudes and fixed frequency (30 Hz). ({\bf C}) Angle between the trajectory and the director ($\theta$) decreases linearly with field at a slope of $-43.2\pm2.3^{\circ} \upmu$mV$^{-1}$. ({\bf D}) Change of direction of motion of a  particle by altering the frequency recursively between 30 Hz (purple) to 50 Hz (yellow), keeping the field amplitude constant at $1.5$ V$\upmu$m$^{-1}$. (Movie S3~\cite{sup}]. ({\bf E}) Time coded trajectories of a particle at  different frequencies and  fixed amplitude (1.5 V$\upmu$m$^{-1}$). ({\bf F}) $\theta$ increases linearly with $f$ at a slope of $1\pm0.05^{\circ}$ Hz$^{-1}$. By ``relative coordinate'' we mean relative to the starting point of each trajectory. Error bars represent the standard deviation of the mean value. Cell thickness: 5.2$\upmu$m.}
\label{fig:figure4}
\end{figure*}

For Fig.\ref{fig:figure2}, we consider a local cartesian coordinate system and small angle approximation, ${\bf \hat{n}}_0 \simeq \hat{\bf x}$ and $\delta {\bf \hat{n}} \sim (0,\theta(x,z))$. The second term of the \eqref{forcelinear} reduces to  $\boldsymbol{\mathcal{F}_{II}}\simeq 2\epsilon_0 (-\Delta \epsilon+ \Delta \sigma  G^{-1}_{\epsilon} G_{\sigma} )\theta \partial_{x} \partial_{z}\Psi_0{\bf E}_0 $, with signs $(- + - +)$ {for positive $z$ and $(+ - + -)$ for negative $z$, as one moves from positive $x$ towards negative $x$, and hence has only} a higher multipole contribution. The force density of \eqref{forceden}, on the other hand, reads,
\begin{equation} 
\label{forcedenf}
\boldsymbol{\mathcal{F}}  \simeq \epsilon_0(-\Delta \epsilon+ \Delta \sigma G^{-1}_{\epsilon} G_{\sigma} ) \Big(\dfrac{\partial \theta}{\partial z}\dfrac{\partial \Psi_0}{\partial x}+\dfrac{\partial \theta}{\partial x}\dfrac{\partial \Psi_0}{\partial z}\Big) E_{0}\hat{\bf z} 
\end{equation} 

From Fig.\ref{fig:figure2} we see that the director curvatures in \eqref{forcedenf} are composed of bend concentrated just outside the Saturn ring ($\partial_{x} \theta$) coupled to ($\partial_{z} \Psi_0$) and splay on the particle surface ($\partial_{z} \theta$ ) coupled to ($\partial_{x} \Psi_0$). We see the signs in four quadrants are $\partial_{x} \theta$ $(- - + +)$, $\partial_{z} \Psi_0$ $(- - - -)$ $\partial_{z} \theta$  $(+ - - +)$, $\partial_{x} \Psi_0$ $(- + - +)$.  From \eqref{Gsigma} and \eqref{Gepsilon} we see that  although $G^{-1}_{\epsilon} G_{\sigma}$ has a nonlocal piece decaying as $1/r^3$, is formally a positive operator if examined in Fourier space. For our system $\Delta \epsilon < 0$ and $\Delta \sigma > 0$. Therefore, the splay contribution produces a force dipole of contractile or puller type while the bend produces a force dipole of extensile or pusher type with respect to the electric field axis.

 Given that the electric field, via \eqref{forcedenf}, results in a force density along $z$, we can understand trajectory III and VII (see Fig.\ref{fig:figure1}D) by asking how the Janus character breaks symmetry in the $xz$ plane. In this plane the splay contribution is greater than the bend as it is present over a larger part of the particle surface. Shifting the force dipole towards the metallic side yields a puller type force dipole resulting in motility with the dielectric face forward as shown in Fig.\ref{fig:figure3}A. To test whether this idea makes sense let us apply it to the case of trajectories I and V of Fig.\ref{fig:figure1}D. Here the breaking of symmetry in the $yz$ plane is of relevance. Bend all along the Saturn ring is in play, which gives a pusher force dipole. Shifting this towards the metal face results in motility with the conductor face forward as shown in Fig.\ref{fig:figure3}B. The consistency of our explanation for the cases of Janus axis parallel and perpendicular to the macroscopic nematic alignment is reassuring. For other trajectories motility is due to a combination of both the effects stated above and therefore interpolates in direction.\\


When the Janus vector ${\bf \hat{s}}$ is oriented neither parallel nor perpendicular to ${\bf \hat n}$, their direction of motion can be controlled by changing the amplitude and frequency of the  field as shown in Fig.\ref{fig:figure4}, A and D, respectively. The moving direction is changed recursively at different points (Fig.\ref{fig:figure4}A) by altering the field amplitude between 1.5 V$\upmu$m$^{-1}$ to 2.0 V$\upmu$m$^{-1}$ at a fixed frequency (Movie S2~\cite{sup}). Figure \ref{fig:figure4}B shows the trajectories of a Janus particle at different fields for the fixed orientation of the metal-dielectric interface. The angle $\theta$ their velocity makes with the director $\bf{\hat n}$ decreases with increasing field (Fig.\ref{fig:figure4}C), with a linear dependence over the range explored. Figure \ref{fig:figure4}D shows the variation in direction of motion along a trajectory as frequency is changed from 30 Hz to 50 Hz (Movie S3 ~\cite{sup}) at a fixed field amplitude. Figure \ref{fig:figure4}E shows the trajectories of a particle at various frequencies in which $\theta$ increases linearly (Fig.\ref{fig:figure4}F). 

\begin{figure}[!h]
\begin{center}
\includegraphics[scale=0.4]{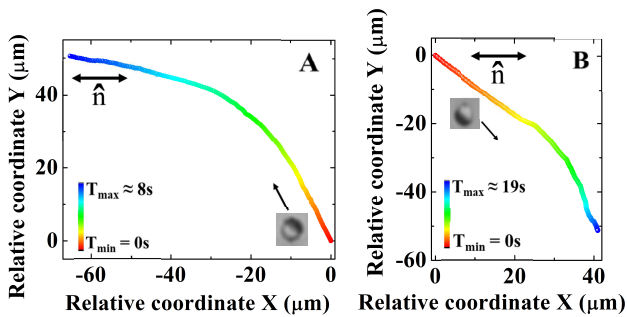}
\end{center}
\caption{({\bf A}) A particle with $\varphi=135^{\circ}$ is piloted to a predetermined place by changing the amplitude  of the field (see Movie S4~\cite{sup}). The field amplitude is increased in a continuous manner from 1.2 V$\upmu$m$^{-1}$ to 2.0 V$\upmu$m$^{-1}$, keeping the frequency fixed at 30 Hz. ({\bf B}) A particle with $\varphi=315^{\circ}$ is transported to a predetermined place by increasing the frequency in a continuous manner from 20 to 90 Hz, keeping the amplitude fixed at 1.6 V$\upmu$m$^{-1}$( see Movie S5~\cite{sup}). The arrows near the particles show the initial direction of motion. Cell thickness: 5.2$\upmu$m.}
\label{fig:figure5}
\end{figure}

Particles with appropriately chosen orientation of the Janus vector can be transported to any predetermined place in the plane of the sample by varying the amplitude and  frequency of the AC field. Figure \ref{fig:figure5}A shows the trajectory of a  particle whose Janus vector ${\bf\hat s}$ is oriented at an angle $\varphi=135^{\circ}$ with respect to ${\bf\hat n}$ (Fig.S2C, Supplementary Material~\cite{sup}). The particle is piloted to a predetermined location on the left with respect to the starting point by increasing the amplitude of the field  at a fixed frequency. Similarly, a particle with $\varphi=315^{\circ}$ is guided to a specified destination by increasing the frequency, keeping the amplitude fixed (see Fig.\ref{fig:figure5}B). In both cases the direction of motion changes continuously while the orientation of the Janus vector ${\bf\hat s}$ remains unchanged (see Movies S4 and S5)~\cite{sup}.

\begin{figure}[!h]
\begin{center}
\includegraphics[scale=0.4]{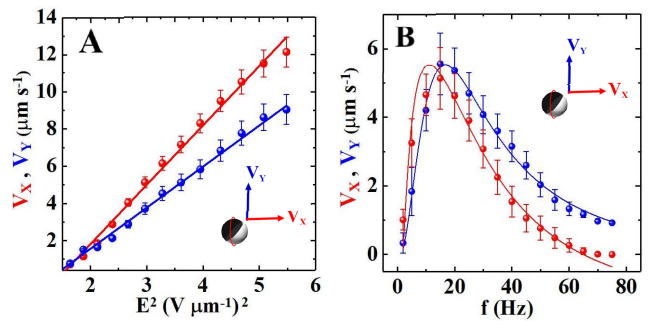}
\end{center}
\caption{({\bf A}) Dependence of velocity components $V_{x}$ and $V_{y}$ of a Janus particle (trajectory-II, Fig.\ref{fig:figure1}D) on electric-field amplitude $E$ at $f$=30Hz.  Solid lines show least-squares fit to $V\propto E^{2}$. Slopes of $V_{x}$ and $V_{y}$ are $3.2$ $\upmu$m$^{3}$V$^{-2}$s$^{-1}$ and 2.2~ $\upmu $m$^{3}$V$^{-2}$s$^{-1}$ respectively. ({\bf B}) Frequency dependent ($E=1.54$ V$ \upmu $m$^{-1}$) velocity components $V_{x}$ and $V_{y}$. Red and blue lines show theoretical fit to Eq.(2). Fit parameters are: $\tau_{e}= 0.25$ s, $\tau_{p}= 0.032$ s for $V_{x}$ and $\tau_{e}= 0.08$ s, $\tau_{p}= 0.04$ s for $V_{y}$. Error bars represent the standard deviation of the mean value. Cell thickness: 5.2$\upmu$m.}
\label{fig:figure6}
\end{figure}

 To understand the navigation of the Janus quadrupolar particles, we measured the field dependence of the velocity (see Fig.S3, Supplementary material~\cite{sup}).
The motion of the particles shown in Fig.\ref{fig:figure4}, A and D  have two velocity components, namely $V_{x}$ and $V_{y}$ along the $x$ and $y$ directions respectively, both of which are proportional to $E^2$ but the slope of $V_x$ is larger than $V_y$ (see Fig.\ref{fig:figure6}A).

When the field is increased, the relative enhancements are unequal, i.e., $\Delta V_x$ is larger than $\Delta V_y$, and consequently $\theta=\tan^{-1}(\Delta V_y/\Delta V_x)$ decreases. The velocity components also depend strongly on the frequency ($f$) of the field (see Fig.\ref{fig:figure6}B). Both the components show the following frequency dependence, given by~\cite{baz1,oleg1}
\begin{equation}
V_{i}(\omega) = {V^{o}_{i}} \frac{\omega^{2} \tau^{2}_{e}}{(1+\omega^{2} \tau^{2}_{p})(1+\omega^{2}\tau^{2}_{e})} 
\end{equation}
where $i=x,y$; $\omega=2\pi f$ is the angular frequency of the applied field, $\tau_{e}$ and $\tau_{p}$ are the characteristic electrode and particle charging time, respectively. Both the velocity components increase as $f^{2}$ in the low frequency regime but decrease as $f^{-2}$ at higher frequencies because the ions cannot follow the rapidly changing field. 
For frequencies above 15 Hz both $V_{x}$ and $V_{y}$ are proportional to $1/f^{2}$ but the coefficient of the decrease of $V_{x}$ is larger than that for $V_{y}$, resulting in an increase in the angle $\theta$ when $f$ is increased. 

Two further remarks are in order. One, for ICEO in an isotropic medium the flows are always of puller type with respect to the electric field axis~\cite{SquiresBazantJFM2006}, ruling out directional control of the type we discuss. Second, we expect that hydrodynamic torques arising from the coupling between the squirmer flow field and anisotropic viscosity of the NLC~\cite{lint} cannot operate here as the orientation of the force dipole driving our swimming particles is determined by the direction of the imposed electric field. This rules out spontaneous change of direction of motion of particles due to such torques at a fixed field and frequency.\\


We have shown that metal-dielectric Janus particles in a nematic liquid crystal film subjected to a perpendicular AC electric field behave like steerable active particles whose direction of motion can be dictated purely by varying the field amplitude and frequency. The underlying mechanism involves the contrasting electrostatic boundary conditions on the two Janus faces of the particles, the dielectric anisotropy of the nematic, and anchoring on the particle surfaces. We show that the time-averaged electrostatic force density produced around the Janus particle by the AC field is that of a force dipole whose center is shifted towards the conducting face, causing the particle to swim in the plane transverse to the field, with dielectric (metal) face forward for particle axis parallel (perpendicular) to the nematic director and interpolating smoothly for intermediate orientations. Studies on the motility at higher concentrations, as well as collective  dynamics as for diffusiophoretic active colloids~\cite{saha1,saha2}, are natural experimental and theoretical challenges. Our study has focused on spherical particles in nematics with perpendicular surface anchoring and linear macroscopic alignment, but we have preliminary results for particles with planar surface anchoring, as well as for race-track director configuration. The abundance of new particles with controlled shapes, surface anchoring~\cite{ivan3} and genus~\cite{ivan2} now available, and their extraordinary topological~\cite{ivan1} and dynamical properties~\cite{sag1,sag2} promise a wide range of as yet unexplored physical effects and their applications.

\vskip 0.2cm
*Corresponding author: sdsp@uohyd.ernet.in\\
\newline
{\bf ACKNOWLEDGEMENTS:}
 SD thanks Steve Granick for hosting his visit to IBS, UNIST which resulted in very useful discussions. SD  also acknowledges Myeonggon Park and Joonwoo Jeong of UNIST for various discussions. We thank O. D. Lavrentovich, I. Mu\v{s}evi\v{c}, S. Bhattacharya and P. Anantha Lakshmi for useful discussions. We also thank K.V. Raman for help in preparing Janus particles.\\
 \newline
  {\bf Funding:} This work is supported by the DST, Govt. of India (DST/SJF/PSA-02/2014-2015). SD acknowledges a  Swarnajayanti Fellowship and DKS an INSPIRE Fellowship from the DST. SR was supported by a J. C. Bose Fellowship of the SERB, India and by the Tata Education and Development Trust.

\end{document}


\fontsize{12}{16}\selectfont
\begin{center}
{\Large{Supplementary materials for}}
\end{center}

\vspace{0.07 cm}
\begin{center}
\textbf{\Large{Omnidirectional transport and navigation of Janus particles through a nematic liquid crystal film}}

\vspace{0.4 cm}
Dinesh Kumar Sahu$^{1}$, Swapnil Kole$^{2}$, Sriram Ramaswamy$^2$, Surajit Dhara$^{1\ast}$

$^{1}$School of Physics, University of Hyderabad, Hyderabad 500 046, India

$^{2}$Centre for Condensed Matter Theory, Department of Physics, Indian Institute of Science\\
 Bangalore 560 012, India

\end{center}

\vspace{0.1 in}
\noindent $^{*}$To whom correspondence should be addressed: sdsp@uohyd.ernet.in\\

\vspace{0.1 in}
\noindent \textbf{This PDF file includes:}\\\\
Supporting Experimental Results\\
Figs. S1 to S3\\
Captions for Movies S1 to S5\\\\\
\\


\noindent \textbf{I. Supporting experimental results}

\noindent \underline{Particle fabrication}\\

\begin{figure}[!ht]
\begin{center}
\includegraphics[scale=0.6]{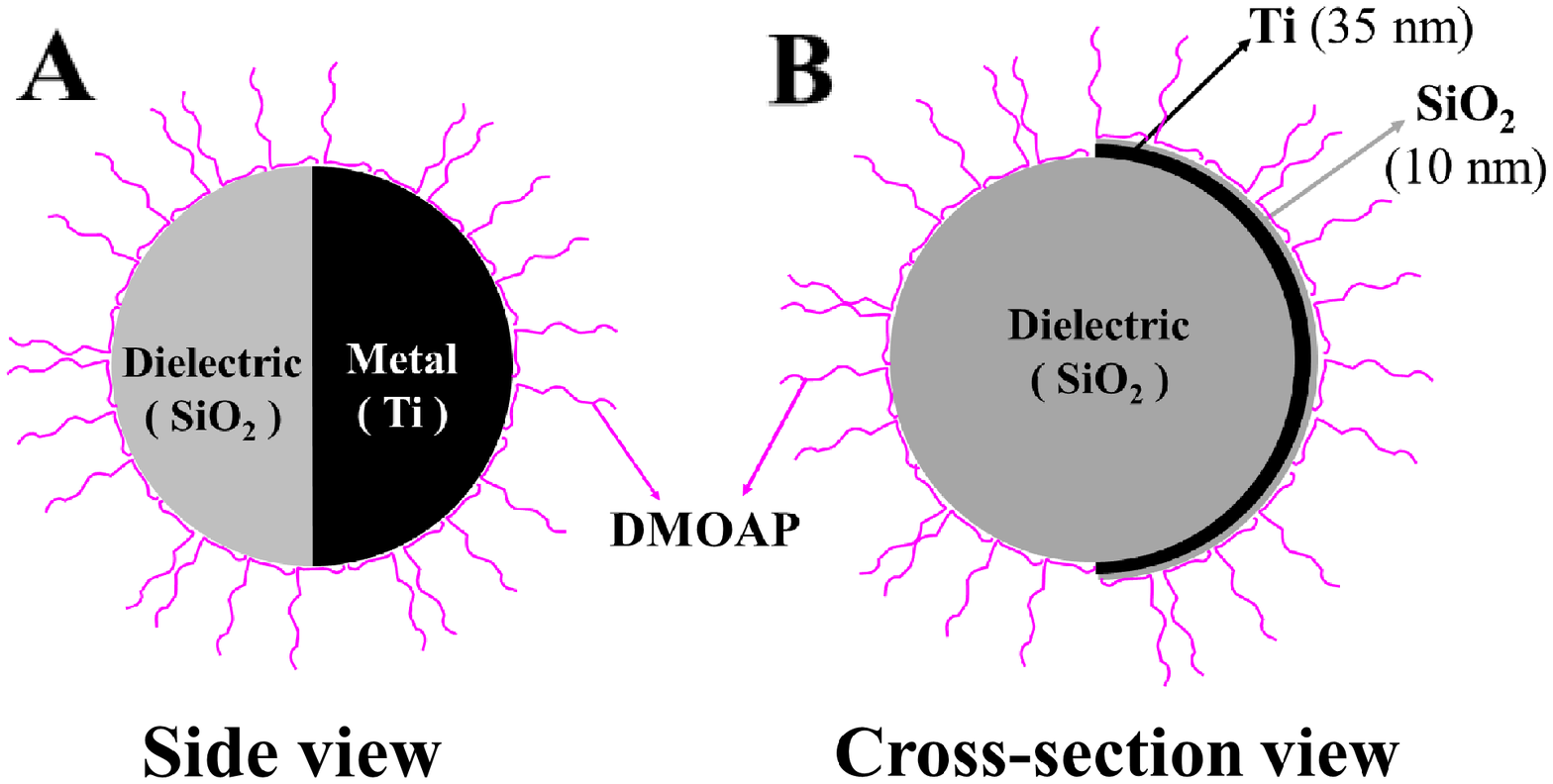}
\end{center}
\textbf{Fig.S1.}({\bf A}) Top and ({\bf B}) side view of a metal-dielectric Janus particle prepared by coating 35nm Ti layer on SiO$_2$ microparticle. On top of Ti, 10nm of SiO$_2$ is coated. Magenta colour strands represent DMOAP molecules.
\label{fig:figure1}
\end{figure}
Metallo-dielectric Janus particles are prepared using directional deposition of metal onto dry silica particles (SiO$_{2}$) of diameter $2a = 3.0 \pm 0.2$ $\upmu$m (Bangs Laboratories, USA) in vacuum [4]. Approximately $2 \% $ suspension ($25$ $\upmu$l) of silica particles is spread on a half glass side ($40$ mm$\times25$ mm), which is pretreated with Piranha solution and dried to form monolayer. Next, a thin Titanium layer of thickness $35$ nm is deposited vertically using electron-beam deposition, at a pressure of $3\times 10^{-6}$ torr, and deposition rate of 0.5 {\AA}s\textsuperscript{-1}. On top of this, a thin layer of SiO$_{2}$ film (10-15 nm) is deposited. Then the slides with silica monolayer are washed thoroughly with deionized water (DI) and isopropyl alcohol. The particles are detached from the slides by ultrasonication in deionized water of volume $20$ ml for $100$ s. Further, the collected particles are sonicated for $30$ minutes to breakup any agglomeration of the Janus particles. After the sedimentation of Janus particles at the bottom of the centrifuge tube, the concentrated suspension is used for the next step. The surface of the Janus particles is coated with N, N-dimetyl-N-octadecyl-$3$  aminopropyl-trimethoxysilyl chloride (DMOAP) in order to induce perpendicular (homeotropic) orientation of the liquid crystal director [41-43]. A schematic diagram showing successive coatings is presented in Fig.S1.

A small quantity (1 wt$\%$) of DMOAP coated Janus particles is dispersed in nematic liquid crystal, MLC-6608.  The liquid crystal obtained from Merck is used directly without any further purification. It exhibits the following phase transitions: SmA $-30^{0}$C N $90^{0}$C Iso. The dielectric anisotropy of MLC-6608 is negative ($ \Delta\epsilon = \epsilon_{\parallel} - \epsilon_{\perp} = -4.2\pm0.1$, where $\epsilon_{||}$ and $\epsilon_{\perp}$ are the dielectric permittivities for electric field {\bf E}, parallel and perpendicular to ${\bf\hat n}$)
whereas the conductivity anisotropy is positive ($ \Delta\sigma =\sigma_{\parallel}-\sigma_{\perp} \simeq 6\times10^{-10}$ Sm\textsuperscript{-1} at $100$ Hz). There is no electroconvection observed in the experimental field and frequency range.

An inverted polarising optical microscope (Nikon Ti-U) with water immersion objective (Nikon, NIR Apo $60/1.0$) is used for observing the particles. A laser tweezer is built on the microscope using a cw solid-state laser operating at $1064$ nm (Aresis, Tweez $250$si). An acousto-optic deflector (AOD) interfaced with a computer is used for trap movement.  A charge-coupled device (CCD) video camera (iDs-UI) at a rate of 50-100 frames per second is used for video recording of the particle trajectory. A particle tracking program is used off-line to track the centres of the particles, with an accuracy of $\pm 10$ nm. 

 The orientation of the Janus vector ${\bf\hat{s}}$ is changed with the help of the laser tweezers through a photo-thermal effect [44,45].  In this method, a particle and its surroundings are locally heated by laser light. At this condition, the trapping point is rapidly moved in the sample plane and instantly the laser is switched off. Consequently, a deformed director field appears around the particle in a fraction of a second. The particle is displaced and also tends to rotate when the director field attains an equilibrium configuration after the laser is switched off. This procedure is repeated  to get a desired orientation of the Janus vector  ${\bf\hat{s}}$ with respect to the director ${\bf\hat{n}}$. \\

\noindent \underline{Omnidirectional transport of quadrupolar Janus particles}:

Figure S2A shows the optical microscope texture of a few Janus quadrupolar particles without cross polarisers. The metal hemisphere (dark) of particles is oriented in different directions, always keeping the Saturn rings perpendicular to the macroscopic director.
The direction of motion of the particles in the sample plane under AC field depends on the orientation of the metal-dielectric interface with respect to the nematic director ${\bf \hat{n}}$. Figure S2B shows the trajectories of several particles with different orientations of the metal-dielectric interface, which is summarised in Fig.S2C. The orientation of the Janus vector ${\bf\hat{s}}$ and hence the angle $\varphi$ is varied from $0^{\circ}$ to $360^{\circ}$ with the help of the laser tweezers as described in the main manuscript. \\



\begin{figure*}[!ht]
\begin{center}
\includegraphics[scale=0.7]{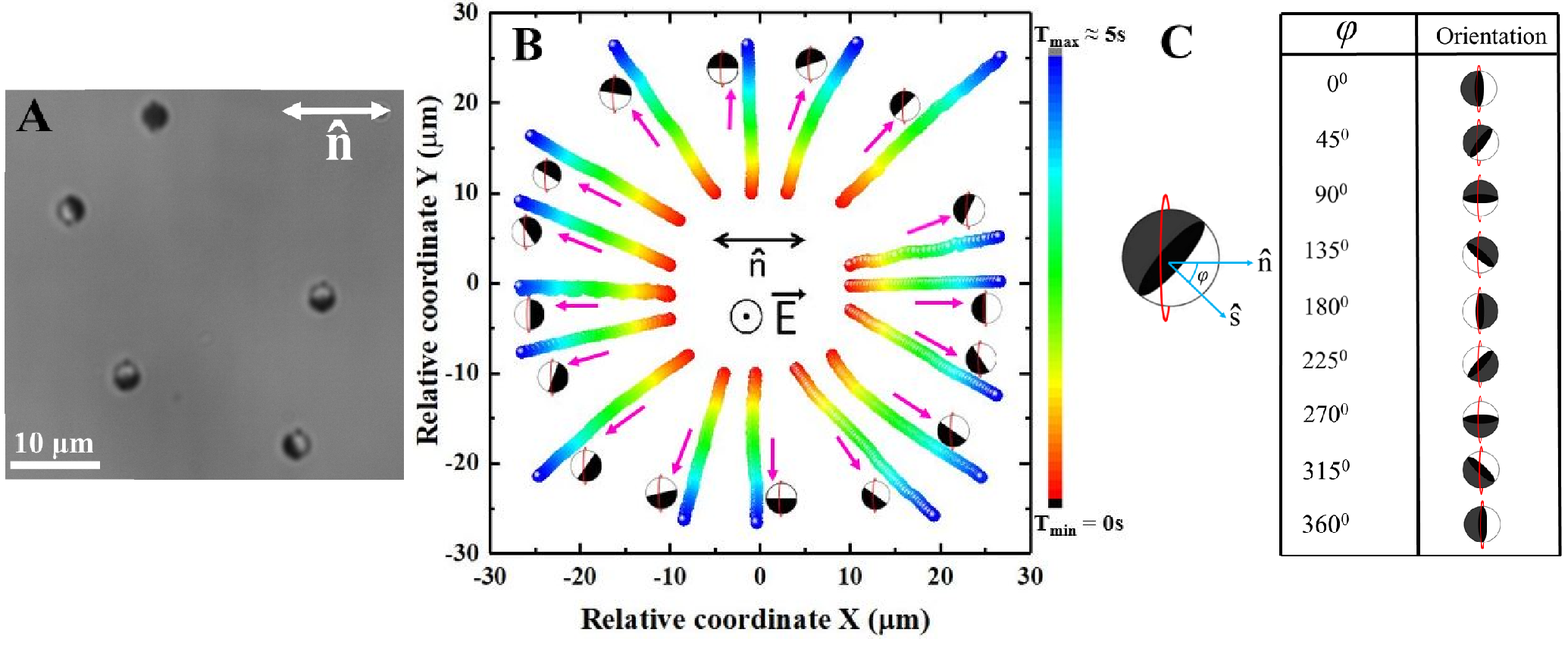}
\end{center}
\textbf{Fig.S2.}({\bf A}) Optical microscope texture of a few Janus quadrupolar particles in the nematic liquid crystal (without polarisers).  Metal hemisphere appears dark under transmitted light.  ({\bf B}) Time coded trajectories of particles in many directions. Directions of ${\bf\hat{n}}$ and {\bf E} ($f=30$ Hz, $E=1.54$ V$\upmu$m$^{-1}$) are shown in the central region. Magenta arrows indicate the direction of motion of the particles. By ``relative coordinate'' we mean relative to the starting point of each trajectory. Cell thickness: 5.2$\upmu$m. ({\bf C}) $\varphi$ is the angle between the director ${\bf\hat{n}}$ and Janus axis ${\bf\hat{s}}$ (normal to the metal-dielectric interface). Selected orientation of a particle in the $xy$ plane at different $\varphi$.
\end{figure*}

\noindent \underline{Electric field dependent velocity of quadrupolar Janus particles}:

Figure S3A, B and C show the electric field dependent velocity of three Janus particles in trajectories-III, I and II, respectively (see Fig.1D).  In all these trajectories $V\propto E^{2}$, with their respective slopes given by $3.5\pm0.1$ $\upmu $m$^{3}$V$^{-2}$s$^{-1},\hspace{0.1cm} 2.1\pm0.2$ $\upmu $m$^{3}$V$^{-2}$s$^{-1}$ and $3.6\pm0.1$ $\upmu $m$^{3}$V$^{-2}$s$^{-1}$.  The slope of $V_{x}$ is greater than that of $V_{y}$ as the particles can move easily along the director rather than in the perpendicular direction. The velocity components, $V_{x}$ and $V_{y}$ of $V_r$ (trajectory-II) are shown in Fig.6A in the main manuscript.\\\\

\begin{figure*}[!ht]
\begin{center}
\includegraphics[scale=0.8]{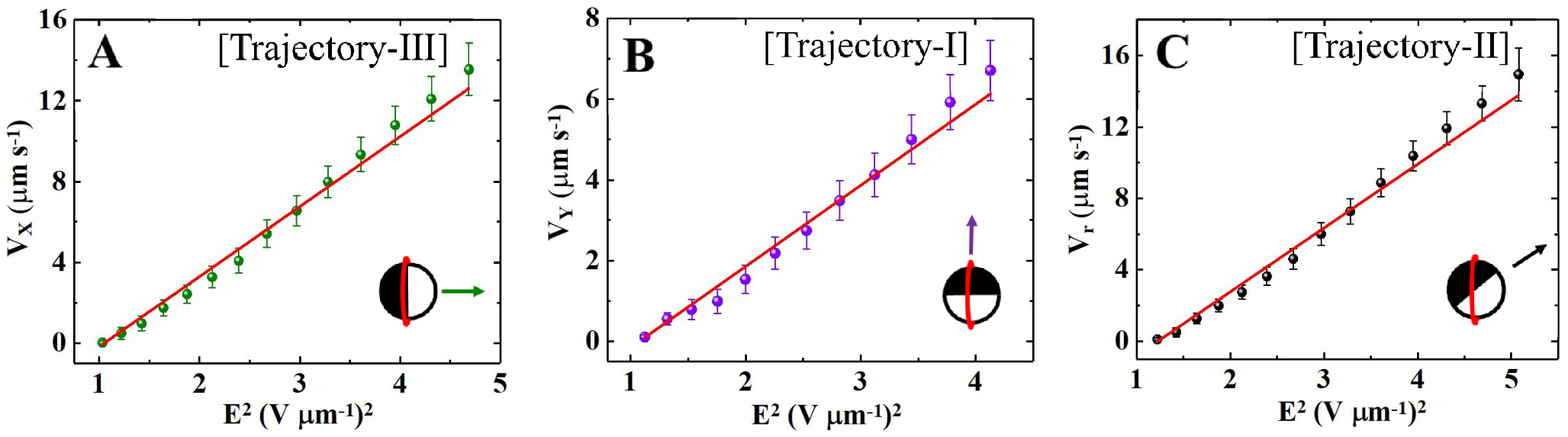}
\end{center}
\textbf{Fig.S3.} Electric field dependent velocity of particles in ({\bf A}) trajectory-III ($V_{x}$), ({\bf B}) trajectory-I ($V_{y}$) and ({\bf C}) trajectory-II ($V_{r}$) (see Fig.1D). Arrows near the spheres indicate the direction of motion. Solid lines show least squares fit to $V\propto E^{2}$. Frequency = 30 Hz. Error bars represent the standard deviation of the mean value. Cell thickness: 5.2$\upmu$m.
\end{figure*}

\newpage
\noindent \textbf{II. Movies}\\

\noindent {\textbf{Movie S1}: Omnidirectional transport of quadrupolar Janus particles. Particles moving along different directions in the plane of the cell due to applied sinusoidal AC field ($1.54$ V$\upmu $m$^{-1}$, $30$ Hz). Trajectories of selected particles are grouped. Metal hemisphere appears darker. The direction of electric field, {\bf E} and the director (${\bf\hat n}$) for all trajectories are shown at the centre. Cell thickness: 5.2$\upmu$m. }\\

\noindent {\textbf{Movie S2}: 
Controlling the direction of motion of a quadrupolar Janus particle by changing the field strength recursively between $1.5$ V$\upmu $m$^{-1}$ to $2.0$ V$\upmu$m$^{-1}$, keeping the frequency constant at $30$ Hz.\\

\noindent {\textbf{Movie S3}: Controlling the direction of motion of a quadrupolar Janus particle by changing the frequency recursively between $30$ Hz to $50$ Hz, keeping the field constant at $1.5$ V$\upmu$m$^{-1}$.\\

\noindent {\textbf{Movie S4}: Controlled transport of a quadrupolar Janus particle by continuously increasing the field amplitude from 1.2 V$\upmu$m$^{-1}$ to 2.0 V$\upmu$m$^{-1}$ at a fixed frequency of 30 Hz. The director (${\bf\hat n}$) is shown at the upper corner of the right hand side.  \\

\noindent {\textbf{Movie S5}: Controlled transport of a quadrupolar Janus particle by continuously increasing the frequency of the field from 20 to 90 Hz at a field amplitude of $1.5$ V$\upmu$m$^{-1}$. The director (${\bf\hat n}$) is shown at the upper corner of the right hand side. \\


\noindent \textbf{References:}\\

\noindent[41] I. Mu\v{s}evi\v{c}, Liq. Cryst. Today \textbf{19}, 2 (2010).

\noindent[42] U. Tkalec, M. Ravnik, S. \v{Z}umer, and I. Mu\v{s}evi\v{c}, Phys. Rev. Lett. \textbf{103}, 127801 (2009).

\noindent[43] M. \v{S}karabot, M. Ravnik, D. Babi\v{c}, N. Osterman, I. Poberaj, S. \v{Z}umer, I. Mu\v{s}evi\v{c}, A. Nych, U. Ognysta, and  V. Nazarenko, Phys. Rev. E \textbf{73}, 021705 (2006).

\noindent[44] M. \v{S}karabot, M. Ravnik, S. \v{Z}umer, U. Tkalec, D. Babi\v{c}, N. Osterman and  I. Mu\v{s}evi\v{c},  Phys. Rev. E \textbf{77}, 031705 (2008). 

\noindent[45] U. Ognysta, A. Nych, V. Nazareko, M. \v{S}karabot and  I. Mu\v{s}evi\v{c}, Langmuir {\bf 25}(20), 12092 (2009).